\documentclass[aps,prb,twocolumn,showpacs,10pt]{revtex4-1}
\usepackage{graphicx}
\usepackage{xcolor}
\usepackage{bm}

\begin{document}
\title{Orbital-Free Density-Functional Theory Simulations of Displacement Cascade in Aluminum}

\author{Ruizhi Qiu}
\email{qiuruizhi@itp.ac.cn}
\affiliation{Science and Technology on Surface Physics and Chemistry Laboratory, Mianyang 621908, Sichuan, China}
\date{\today}
\begin{abstract}
Here, we report orbital-free density-functional theory (OF DFT) molecular dynamics simulations of the displacement cascade in aluminum. The electronic effect is our main concern. The displacement threshold energies are calculated using OF DFT and classical molecular dynamics (MD) and the comparison reveals the role of charge bridge. Compared to MD simulation, the displacement spike from OF DFT has a lower peak and shorter duration time, which is attributed to the effect of electronic damping. The charge density profiles clearly display the existence of depleted zones, vacancy and interstitial clusters. And it is found that the energy exchanges between ions and electrons are mainly contributed by the kinetic energies.
\end{abstract}
\maketitle

\section{Introduction}

The effects of particle irradiation on the properties of materials have been recognized and studied for over 70 years~\cite{Wigner1942}. In order to predict the consequences of irradiation, it is necessary to understand the atomic-level processes that underlie these effects. The primary source of radiation damage during the particle irradiation is the displacement cascades, i.e., a chain of atomic collisions after a high-energy particle (with a kinetic energy of more than $\sim$1 keV) collides with a host atom. Now the displacement cascades, as well as the subsequent dynamics of point defects, are routinely studied by classical molecular dynamics (MD) simulation with empirical potentials~\cite{Bacon1997,Nordlund1999,Osetsky2003,Bacon2003,Bai2010}. These simulations have provided important insights into the fundamentals of radiation damage, and dramatically advance the understanding of defects and defect processes in a number of materials. However, electronic effects, such as electronic damping, inelastic scattering and thermal conductivity, are often neglected or include only implicitly~\cite{Race2010}.

It is expected that full role played by electrons in displacement cascade will be revealed only by their explicit treatment~\cite{Race2010}. For example, the threshold effect for the electronic stopping power of LiF~\cite{Auth1998,Draxler2005,Markin2009} is explained in the framework of time-independent density-functional theory (DFT)~\cite{Runge1984} and attributed to the existence of the band gap~\cite{Pruneda2007}. Moreover, time-independent Kohn-Sham DFT (KS DFT)~\cite{Hohenberg1964,Kohn1965} simulations of low-energy recoil events in SiC have revealed that significant charge transfer occurs between atoms and influence the displacement threshold energies~\cite{Gao2009}. In addition, the electronic damping and excitation have been investigated by semi-empirical time-dependent tight-binding simulation of model metals~\cite{Mason2007,Race2009}.
However, to our best knowledge, full evolution of a displacement cascade has not been explored due to the small size of simulation cell.
In addition, previous simulation have focused almost exclusively on nonmetals~\cite{Dudarev2013} or model metals.

Orbital-free DFT (OF DFT)~\cite{Hohenberg1964}, which enables the simulation of millions atom system~\cite{Gavini2007,Hung2009}, provides a choice to investigate the full evolution of displacement cascade with quantum mechanics. Unlike KS DFT that uses single-electron orbitals, OF DFT solves directly for the electron density as the sole variable and is significantly less computationally expensive. The accuracy of OF DFT depends on the kinetic energy density-functional (KEDF), which replaces the kinetic energy expression using Kohn-Sham orbitals in KS DFT. State-of-the-art KEDFs, such as Wang-Teter (WT)~\cite{Wang1992} and Wang-Govind-Carter (WGC)~\cite{Wang1999,Wang2001} KEDF, are designed to reproduce the Lindhard linear response of a free-electron gas~\cite{Lindhard1953}. Therefore, OF DFT is limited to simulate the main group, near-free-electron-like metals. However, one should note the advantages of OF DFT compared to the embedded atom method (EAM)~\cite{Daw1983,Daw1984} or Finnis-Sinclair (FS)~\cite{Finnis1984} model which are widely used in the classic MD simulation of metals. First, the electron density is optimized through the self-consistent iteration in each ionic step of OF DFT while it is assumed to be frozen or summation of frozen atomic electron densities in the EAM or FS model. Apparently, more physical insight could be obtained from OF DFT than the classical MD. Second, the interatomic potentials only know the environments in which they were originally fit and lack in transferability, while OF DFT has transferable pseudopotentials. Third, higher accuracy could be expected from the optimization of electron density in the framework of Hohenberg-Kohn theorem compared to classical MD. In fact, OF DFT with WGC KEDF has been proven to be reliable for describing fundamental properties of main group metals with an accuracy comparable to KS DFT~\cite{Shin2009,Chen2013,Shin2014,Sjostrom2014,Das2015,Zhuang2016}. Particularly, the authors systematically calculated formation energies and migration energies of various point defects in aluminum and found that the results from OF DFT are in excellent agreement with those through KS DFT~\cite{Qiu2017}. These validate OF DFT with WGC KEDF as an accurate tool for simulating large-scale systems with defects, such as displacement cascade in this work.

Here we employ OF DFT to simulate and study the displacement cascade in aluminum using a simulation cell of $20\times20\times20$ face-centered-cubic unit cells containing 32000 atoms. This system size enables us to run several thousand MD steps using a modern supercomputer during a reasonable time. In addition, the kinetic energy of primary knock-on atom (PKA) is set as a regular value 1 keV because for higher energy, the resulting displacement cascade will propagate across the periodic boundary and interact with itself, causing unrealistic ion evolution. Nevertheless, our simulation includes the displacement phase, the relaxation phase, and the cooling phase, and thus could directly investigate the electronic effect during the full evolution of displacement cascade. The remainder of this paper is organized as follows. The computational details are described in Section~\ref{sec:method}. The results are presented and discussed in Section~\ref{sec:result}. Section~\ref{sec:conclusion} serves as a conclusion.

\section{Computational methods and details}
\label{sec:method}

The simulations are performed using PROFESS 3.0~\cite{Ho2008,Hung2010,Chen2015}. For calculating the non-interacting kinetic energy, we use WGC non-local KEDF with its three parameters  $\alpha$=$\textstyle{\frac{5}{6}}$+$\textstyle{\frac{\sqrt{5}}{6}}$, $\beta$=$\textstyle{\frac{5}{6}}$-$\textstyle{\frac{\sqrt{5}}{6}}$, and $\gamma$=2.7. The electron-ion interaction is evaluated using bulk-derived local pseudopotential. Local density approximation (LDA)~\cite{Perdew1981} is used to describe the electron exchange and correlation. The kinetic energy cutoff is chosen as 600 eV. Test calculations show that these parameters are reliable for describing the fundamental properties of Al including the formation and migration energies of vacancy and various self-interstitials. To simulate the displacement cascade, the PROFESS code has been modified to include a variable time step with the maximum distance being 0.05 $a_{\rm B}$ with $a_{\rm B}$ being the Bohr radius. The simulated crystal is equilibrated for 1 ps for temperature to be a Gaussian distribution at 100 K, then the primary knock-on atom (PKA) is given kinetic energy to initiate the displacement cascade. The initial velocity direction of PKA is chosen as the three main crystallographic directions, i.e., $\langle100\rangle$, $\langle110\rangle$, and $\langle111\rangle$, and the kinetic energy of PKA is set as 1 keV.

For comparison, we also performed a classical MD simulation of displacement cascade using LAMMPS~\cite{Plimpton1995}. We use the FS interatomic potential developed by Mendelev {\it et al.}~\cite{Mendelev2008} and connect it with the Ziegler-Biersack-Littmark (ZBL) pair potential~\cite{Ziegler1985} to model highly repulsive nuclei-nuclei interaction for small atomic separation. The computational parameters are same as those in OF DFT.

\section{Results and discussion}
\label{sec:result}

\subsection{Displacement threshold energy}

\begin{figure}[t]
\begin{center}
\includegraphics[width=0.49\textwidth]{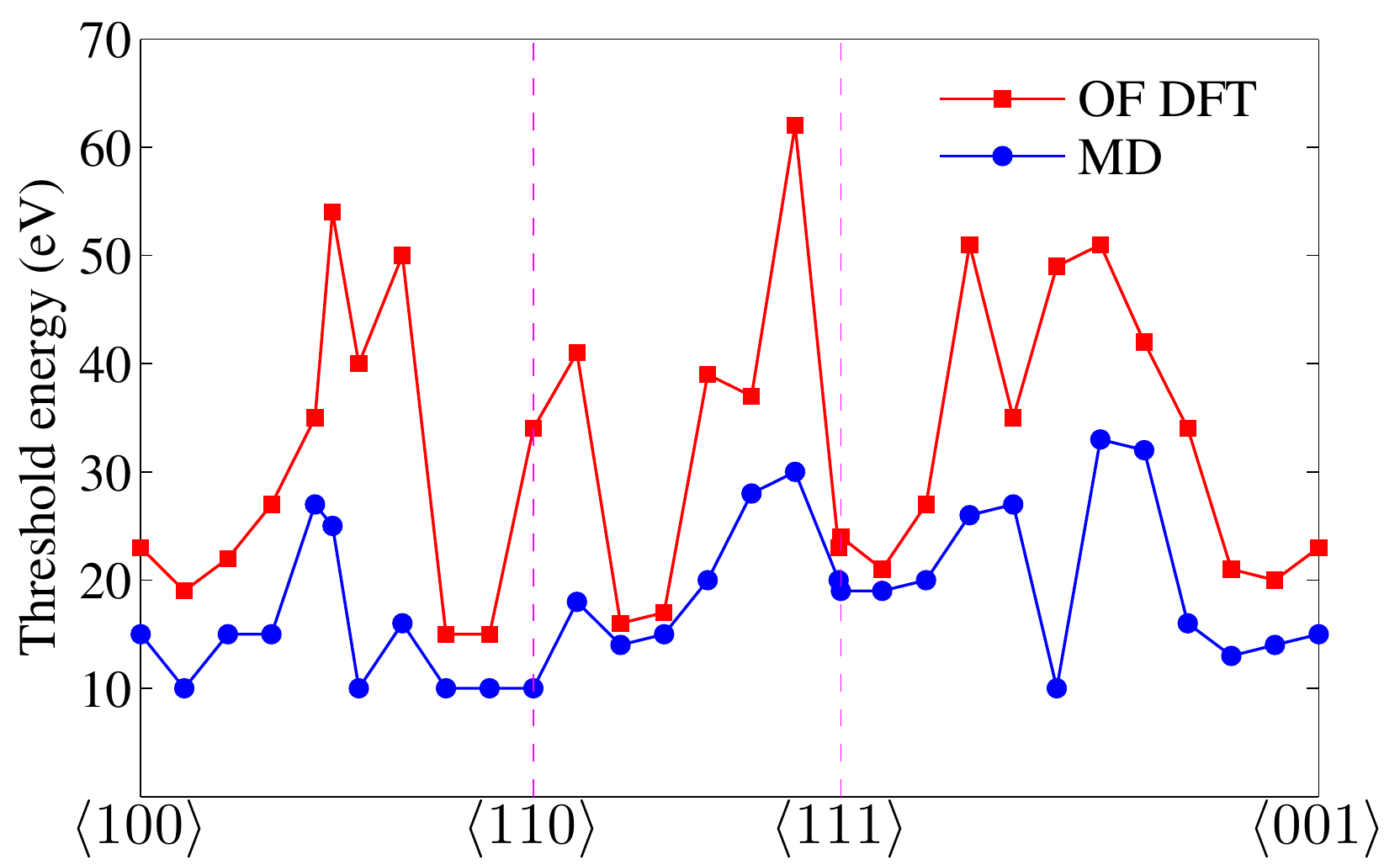}
\caption{Displacement threshold energies of Al calculated using OF DFT and MD along the directions $\langle100\rangle$ $\rightarrow$ $\langle110\rangle$ $\rightarrow$ $\langle111\rangle$ $\rightarrow$ $\langle001\rangle$.}
\label{fig:eth}
\end{center}
\end{figure}

The displacement threshold energy $E_{\rm d}$ is defined as the minimum kinetic energy of PKA necessary to permanently form a stable defect.
It is a key physical parameters being relevant to defect production under irradiation.
Figure~\ref{fig:eth} displayed the calculated value of $E_{\rm d}$ for different direction of PKA.
The results are calculated using OF DFT and MD.
Threshold energy varied considerably depending on the local bonding structures and overall, the curve shape of OF DFT is similar to that of MD.
There are two pockets of low $E_{\rm d}$ around $\langle100\rangle$ and $\langle110\rangle$ surrounded by regions of much higher $E_d$, which is already found in the experiment~\cite{King1983}.
In addition, our calculation show that there is also a shape pocket of low $E_{\rm d}$ around $\langle111\rangle$.

\begin{figure}[hb!]
\begin{center}
\includegraphics[width=0.49\textwidth]{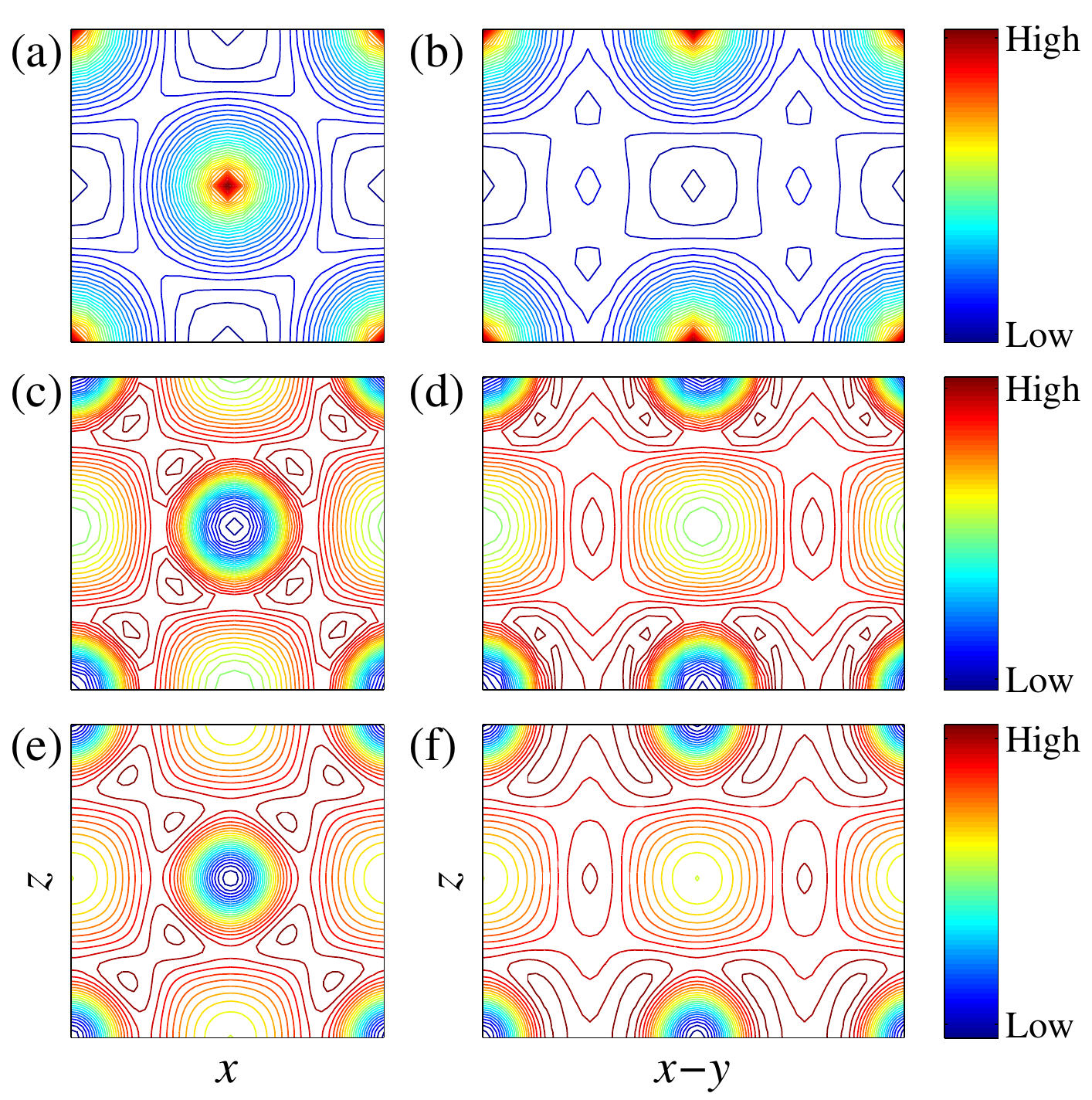}
\caption{Density profile in the $\langle100\rangle$ (a, c, e) and $\langle110\rangle$ plane (b, d, f) from MD (a, b), OF DFT (c, d) and KS DFT (e, f) calculation. Note that the density from OF DFT well reproduced that from KS DFT and differ much from that from MD.}
\label{fig:dens}
\end{center}
\end{figure}

The main differences of $E_{\rm d}$ between OF DFT and MD includes two points.
First, the calculated $E_{\rm d}$ using OF DFT are higher than that from MD. This is attributed to the use of a small lattice parameter in the OF DFT calculation. 
Second, the calculated $E_{\rm d}$ along $\langle110\rangle$ direction from OF DFT is much higher than that from MD.
To understand this, let us note the difference between the charge density of OF DFT and that of MD. 
As can be seen from Figure~\ref{fig:dens}, the charge bridge between Al atoms is present in the OF DFT and KS DFT while absent in MD.
The charge bridges contributes a larger electronic stopping power and then lift the displacement threshold energy along $\langle110\rangle$ direction.
This feature of charge distribution could be incorporated into the improved empirical potentials used in MD.
In addition, since the interaction between ions and electrons is almost determined by the Hartree interaction which is the functional of charge density, the similarity between charge density from OF DFT and that from KS DFT illustrates the validity of OF DFT.

\subsection{Displacement spike}

Now let us turn to the cascade simulation.
Here we will consider the displacement spike, which is the main feature of collision cascade from the point of view of MD.
The displacement spike is initiated by PKA, which moves sufficiently slowly that it interacts with the surrounding ions.
In the simulation, PKA intrude into a undisturbed region which usually taken to be a crystalline lattice for simplicity.
A typical displacement spike could be divided into three phases, i.e., the displacement phase, the relaxation phase and the cooling phase.
The displacement phase is the initial period of the disruption in which many of the interstitial and vacancy defects formed.
There follows the relaxation phase during which the energy is rapidly repartitioned amongst the ions to yield a hot and potentially molten region.
Finally there follows a cooling phase in which the excited region grows and cools.
By the end of the displacement spike, the lattice will heal itself to a large extent, many defects have recombined and the final damage state may form.

\begin{figure}[th]
\begin{center}
\includegraphics[width=0.49\textwidth]{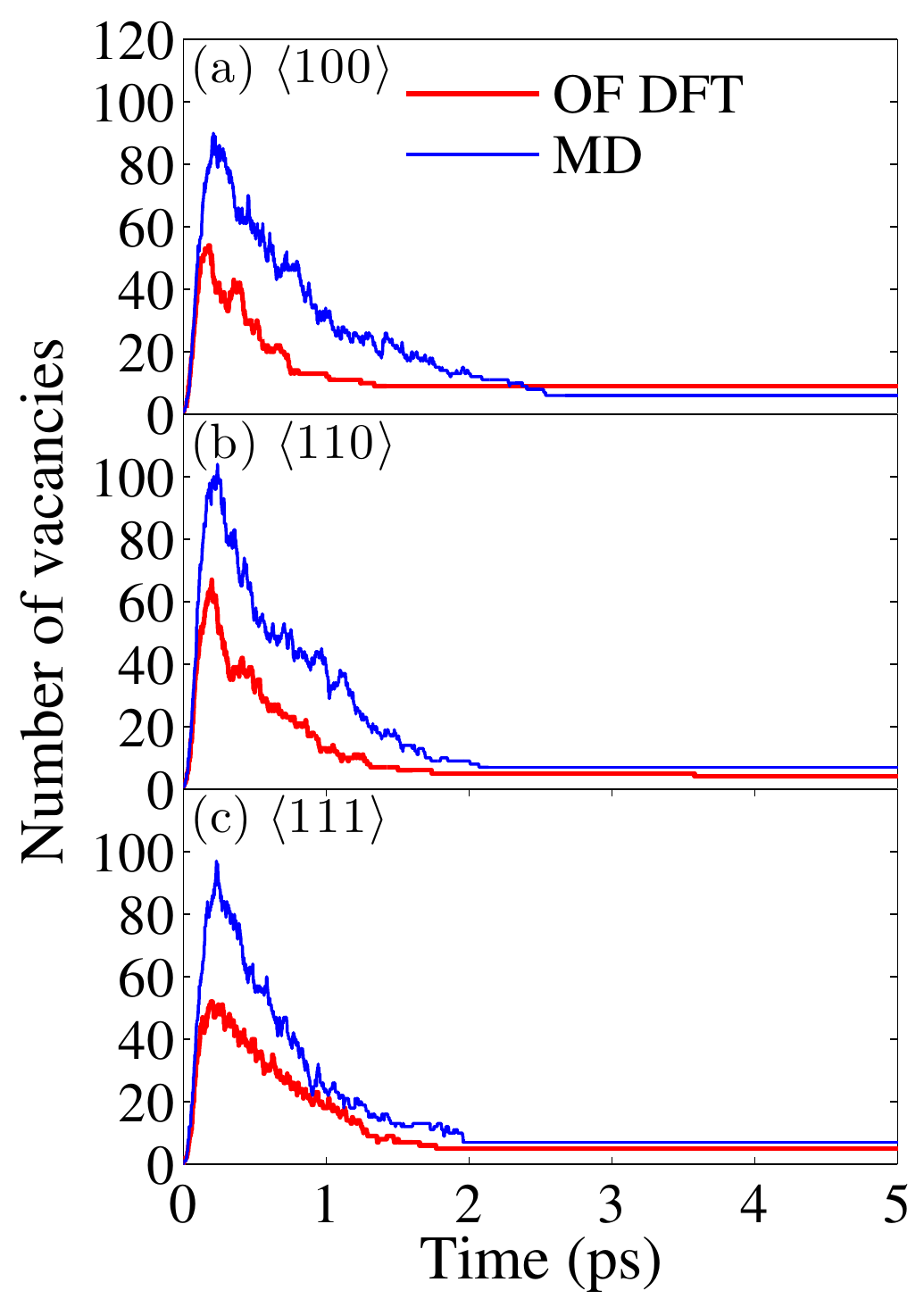}
\caption{Time dependence of number of vacancies in the collosion cascade with PKA energy being 1 keV and PKA direction being three main crystallographic directions.}
\label{fig:ndef}
\end{center}
\end{figure}

To illustrate the displacement spike, we first plot the number of vacancies versus time in Figure~\ref{fig:ndef}.
The number of defects rapidly increase in the $\sim$0.1 ps and then gradually decrease in the following $\sim$2 ps, and finally stepwise decrease.
The three phases of displacement spike are very clear in this figure.
Compared with the results from MD, the peak is lower and the duration time is shorter.
This implies the existence of electronic damping.

\subsection{Electronic effect}

\begin{figure}[th]
\begin{center}
\includegraphics[width=0.49\textwidth]{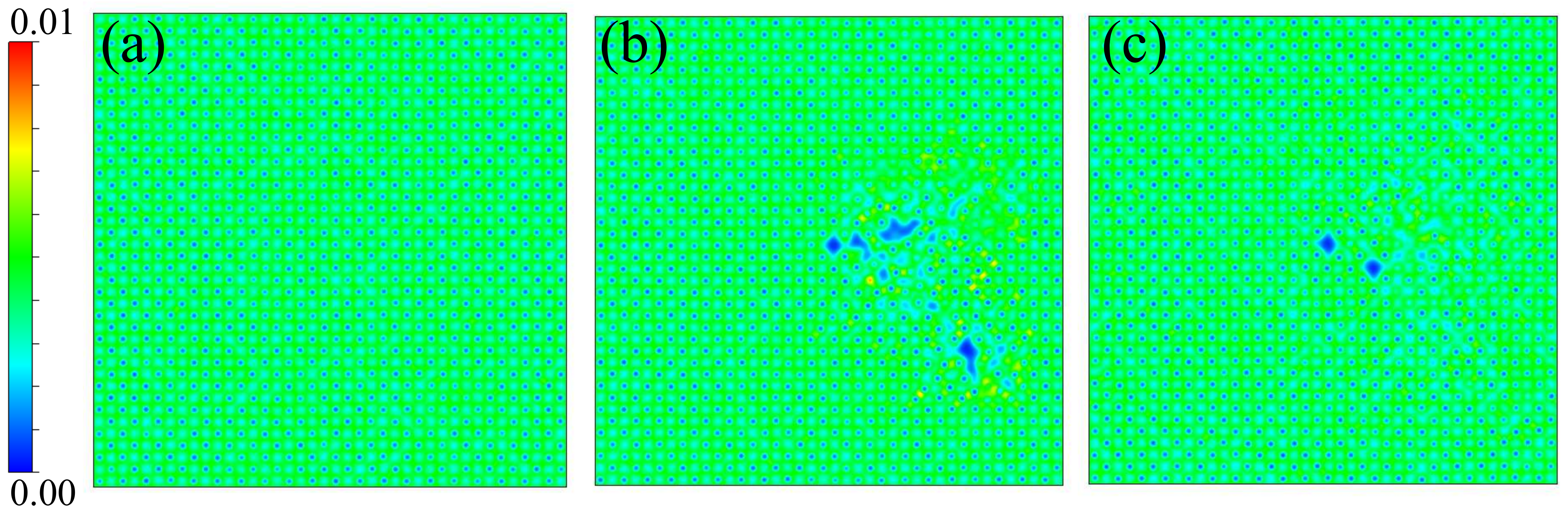}
\caption{The charge density profile of $\langle001\rangle$ plane at the (a) 0.0 ps, (b) 0.1 ps, and (c) 1 ps. Here PKA energy is chosen as 1 keV and the direction is taken along $\langle100\rangle$.}
\label{fig:charge}
\end{center}
\end{figure}

Now OF DFT show its ability to illustrate the electronic effect on the displacement spike.
The electrons are expected to play an important role in the evolution of the displacement spike.
The charge densities of this system at different moments are contoured in Figure~\ref{fig:charge}.
The existence of the depleted zones, vacancy and interstitial clusters is clear.

\begin{figure}[bh]
\begin{center}
\includegraphics[width=0.4\textwidth]{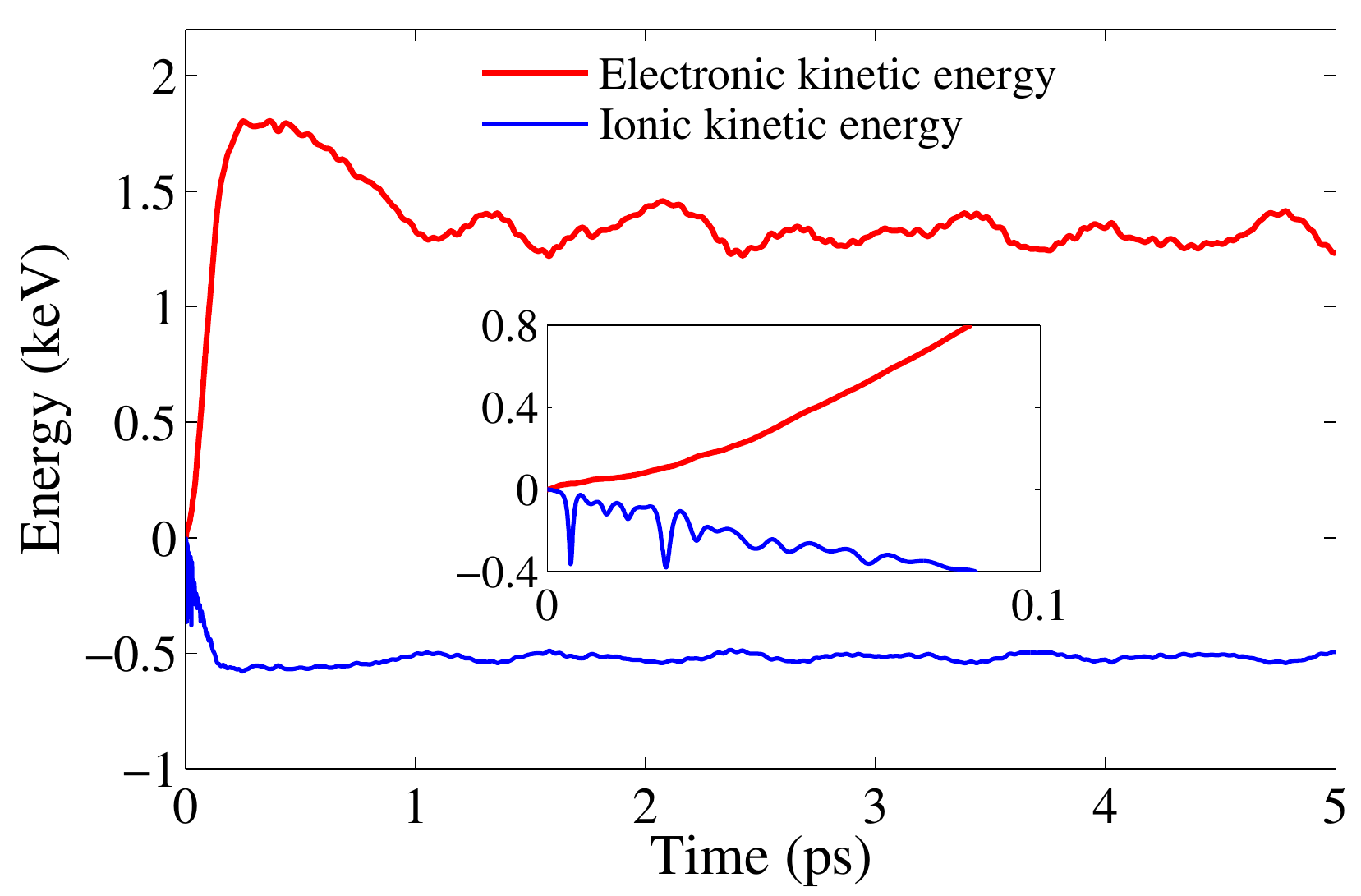}
\caption{Time dependence of ionic and electronic kinetic energy. Here PKA energy is chosen as 1 keV and the direction is taken along $\langle100\rangle$. The inset displays the first 0.1 ps and the dips are ascribed to be the calculation error of the ewald energy when two ions approach very clear.}
\label{fig:eke}
\end{center}
\end{figure}

At last let us turn to the discussion of the energy exchange between ions and electrons, which is shown in the OF DFT simulation and concealed in the MD simulation.
Each term in the expression of the total energy, including kinetic energy density functional, ion-electron interaction energy, Hartree energy, exchange-correlation energy, ion-ion interaction energy, and ionic kinetic energy were examined.
Figure~\ref{fig:eke} plot the ionic kinetic energy and electronic kinetic energy (the calculated value of KEDF) versus time.
There is a net transfer of energy ($\sim$0.5 keV) from the ionic subsystem into the electronic subsystem during the displacement phase.
And during the following phase, the electrons will function as a thermal bath in which the final defect distribution establishes itself.
Clearly, the energy exchanges between ions and electrons are mainly contributed by the kinetic energies.
This could be understood as the simple collision between the electrons and ions.

\section{Conclusion}
\label{sec:conclusion}

In this work, using large-scale orbital-free density-functional theory (OF DFT) simulation of radiation damage in Al, we directly investigate the time evolution of charge density, the electronic damping effect, and the energy exchange between ions and electrons.
The displacement threshold energies are determined for PKA direction along the path $\langle100\rangle$ $\rightarrow$ $\langle110\rangle$ $\rightarrow$ $\langle111\rangle$ $\rightarrow$ $\langle001\rangle$ directions. 
Two pockets of low threshold energy are found around $\langle100\rangle$ and $\langle110\rangle$, which is consistent with the experiment. 
Simulations of displacement cascades with primary knock-on atom energy up to 1 keV are performed using a supercell of 32000 atoms. 
Compared to the classical molecular dynamics simulations, the displacement spike from OF DFT simulation has a lower peak and shorter duration time, which is attributed to the electronic damping. 
The charge density profiles clearly display the existence of depleted zones, vacancy and interstitial clusters. 
And it is found that the energy exchanges between ions and electrons are mainly contributed by the kinetic energies.

\section*{Acknowledgement}
We would like to acknowledge the financial support from National Science Foundation of China (Grant No. 11404299, 21471137, 11504342, and 21601167),
the  Science  Challenge  Project  of  China (Grant No. TZ2016004), the ITER project (Grant No. 2014GB111006), and the CAEP project (Grant No. TCGH0708).

\bibliographystyle{apsrev4-1}
\bibliography{cascade}
\end{document}